# 3D visualizations of nano-scale phase separation and ultrafast dynamic correlation between phases in $(Na_{0.32}K_{0.68})_{0.95}Fe_{1.75}Se_2$

P. C. Cheng[1], W. Y. Tzeng[1], Y. J. Chu[1], C. W. Luo[1,2,3*], A. A. Zhukov[4,5,6*], J. Whittaker[4], J.-Y. Lin[3,7*], K. H. Wu[1], J. Y. Juang[1], M. Liu[8], I. V. Morozov[8,9], A. I. Boltalin[8], A. N. Vasiliev[8,10,11]

[1]*Department of Electrophysics, National Chiao Tung University, Hsinchu 300, Taiwan*

[2]*Taiwan Consortium of Emergent Crystalline Materials, Ministry of Science and Technology, Taipei 10601, Taiwan*

[3]*Center for Emergent Functional Matter Science, National Chiao Tung University, Hsinchu 30010, Taiwan*

[4]*National Graphene Institute, University of Manchester, Manchester, UK*

[5]*Manchester Centre for Mesoscience and Nanotechnology, University of Manchester, Manchester, UK*

[6]*School of Physics and Astronomy, University of Manchester, Manchester, UK*

[7]*Institute of Physics, National Chiao Tung University, Hsinchu 300, Taiwan*

[8]*Lomonosov Moscow State University, Moscow 119991, Russia*

[9]*Lebedev Physical Institute, Russian Academy of Sciences, 119991 Moscow, Russia*

[10]*National University of Science and Technology "MISiS", Moscow 119049, Russia*

[11]*National Research South Ural State University, Chelyabinsk 454080, Russia*

Phase separation of metallic and antiferromagnetic (AFM) insulating phases in alkaline iron selenides ($A_x Fe_{2-y} Se_2$) continues to attract intense interest because the relationship between two peculiar features probably is a key to clarifying the pairing mechanism of $A_x Fe_{2-y} Se_2$ superconductors. Here we report that the 3D visualizations of nano-scale phase separation in $(Na_{0.32}K_{0.68})_{0.95}Fe_{1.75}Se_2$ single crystals are revealed by hybrid focused-ion-beam scanning electron microscopy and the superconducting paths are fully percolative in 3D. Moreover, the phase-related ultrafast dynamics in $(Na_{0.32}K_{0.68})_{0.95}Fe_{1.75}Se_2$ is studied by dual-color pump-probe spectroscopy. The anomalous changes in the electron and acoustic phonon components of transient reflectivity change ($\Delta R/R$) identify two characteristic temperatures $T^*$~100 K (the onset temperature of coupling between nano-metallic and AFM phases) and $T_H$~230 K (the onset temperature of metallic-interface-phase). An energy-transfer channel between the nano-metallic and AFM

**phases is inferred. This proposed channel provides a new insight into the pairing mechanism of alkaline iron selenide superconductors.**

## I. INTRODUCTION

Alkaline iron selenides ($A_x Fe_{2-y}Se_2$) continue to attract intense research interest due to their high critical temperature ($T_c \sim 30$ K) [1–4]. Some particular features, such as the electronic structure, their magnetism and the ordering of the iron vacancies, distinguish $A_x Fe_{2-y}Se_2$ from other iron-based superconductors. For iron-pnictide superconductors, the interband scattering between electron and hole pockets gives rise to electron pairing superconductivity mechanism [5–8], but the absence of hole pockets near the center of the Brillouin zone in $A_x Fe_{2-y}Se_2$ prevents the pairing scenarios proposed in Refs. [9–11]. Spatial phase separation is another intriguing feature associated with the order of iron-vacancies [12,13]. Two main phases in $A_x Fe_{2-y}Se_2$ have been identified: a minority metallic phase and a majority antiferromagnetic (AFM) insulating phase with a large magnetic moment of 3.3 $\mu_B$ per Fe atom [14]. The AFM region with a $\sqrt{5}\times\sqrt{5}$ iron-vacancy order has the stoichiometry of $A_{0.8}Fe_{1.6}Se_2$ (245). The metallic filamentary region meandering within the AFM structure has a composition of $AFe_2Se_2$ (122) and becomes superconducting at the superconducting critical temperature ($T_c$) up to 30K [15–18].

Recently, attempts have been made to find relationship between these two peculiar features because this relation probably is a key to understanding the distinct Fermi-surface topology (the absence of hole pockets near the Brillouin zone center) and clarifying the pairing mechanism of alkali metal iron selenides [19]. The orbital selective Mott phase (OSMP) has been proposed to explain the relationship between the electronic band structure and the characteristics of spatially separated phases [20–22]. The $d_{xy}$ band in $A_x Fe_{2-y}Se_2$ superconductors becomes Mott localized and the $d_{xz}/d_{yz}$ bands remain metallic when the temperature is higher than the OSMP transition temperature ($T_H$). This transition observed in the metallic region of $K_x Fe_{2-y}Se_2$ using angle-resolved photoemission spectroscopy (ARPES) [23] initiates the correlation between the metallic phase and the AFM insulating phase [24]. A recent spatial-resolved micro x-ray diffraction study also demonstrated the existence of an interface phase between the AFM and the superconducting (SC) regions, which is thought to protect electronic coherence of the SC region [22]. However, it remains a matter of debate as to whether the interface phase provides protection for the SC region or it is just a simple link between the AFM and SC phases.

To explore this issue, a hybrid focused ion beam scanning electron microscopy

(FIB-SEM) system has been used to investigate the superconducting/AFM phase separation, and the femtosecond time-resolved spectroscopy has been exploited to study superconductivity [25–33], the coexistence of nano-superconducting/AFM phases, and the mechanism of OSMP at the normal state of $K_xFe_{2-y}Se_2$ single crystals [34].

## II. EXPERIMENTS

In this study, the $(Na_{0.32}K_{0.68})_{0.95}Fe_{1.75}Se_2$ single crystals were grown using a self-flux method [35]. The measurements for the temperature-dependent magnetic moment $M(T)$ were performed using a superconducting quantum interference device (SQUID) from Quantum Design and the onset of the superconducting transition temperature was found to be at about 30 K.

A hybrid FIB-SEM system (Carl Zeiss Crossbeam 540) was used to investigate the 3D mapping [36] of the superconducting phase topology in $(Na_{0.32}K_{0.68})_{0.95}Fe_{1.75}Se_2$ single crystals. The thickness of serially milled off slices was selected in the range of 2.5-9 nm. The SEM images were taken using three different detectors: in-lens, secondary electron and energy dispersive backscattered electrons detector. The best contrast between superconducting and antiferromagnetic components was obtained using the in-lens detector. The sample has been aligned with a- and b-axes employing single crystal surface steps of growth. The area of interest was cut off using three cross-sectional trenches [37]. ImageJ software was employed for three-dimensional (3D) image reconstruction from the generated data stack.

Femtosecond spectroscopy measurements were performed using a dual-color pump-probe system and an avalanche photo-detector, using a standard lock-in technique. For the laser light source, the wavelength and the pulse duration were 800 nm (1.55 eV) and 100 fs, respectively. With the 5.2-MHz repetition rate in our laser light source, the time interval between two pulses was ~192 ns, which is long enough to avoid the energy accumulation effect in $(Na_{0.32}K_{0.68})_{0.95}Fe_{1.75}Se_2$ single crystals. Due to the strong absorption at 400 nm (3.1 eV), the fundamental light of 800 nm was doubled to 400 nm as the pump beam. The polarizations of the pump and probe beams were mutually perpendicular to reduce the background noise, especially from the pump beam, for getting the pump-probe signal with higher signal to noise ratio. The fluences of the pump and probe beams were 30.5 and 3.3 $\mu J/cm^2$, respectively. After excited by the pump pulse with a photon energy of 3.1 eV, the time evolution of the $\Delta R/R$ on a femtosecond timescale was detected using a 1.55-eV femtosecond pulse. Because $(Na_{0.32}K_{0.68})_{0.95}Fe_{1.75}Se_2$ single crystals are easily degraded by water and oxygen at ambient environment, a homemade dry box was used to protect the samples from damp. After the samples were withdrawn from a vacuumed quartz tube, they demonstrated/retained a brilliant crystal surface (proving the good quality of our storage method), and then immediately transferred to a cooling chamber. All of these manipulations were performed inside the homemade dry box.

## III. RESULTS AND DISCUSSION

### A. Morphology of coexisting superconducting and antiferromagnetic phases

Figure 1 shows typical cross sectional images for three main orientations in a $(Na_{0.32}K_{0.68})_{0.95}Fe_{1.75}Se_2$ single crystal obtained from 3D nanotomography results using a hybrid FIB-SEM system Carl Zeiss Crossbeam 540. For more detailed overview of several 3D image stacks can be also found in the supplementary material. As can be seen from Figs. 1(a) and 1(b), narrow (<150nm) quasicontinuous superconducting stripes (areas of brighter contrast) propagate in the direction of $\pm(34\pm3)°$ out of the basal plane. In accordance with the tetragonal (I4mmm) crystal symmetry, the images in Figs. 1(a) and 1(b) are pretty similar.

In the basal plane (see Fig. 1(c)), the superconducting areas demonstrate nearly rectangular shape and appears being arranged into a net of closely connected stripes, which are directed $\pm(45\pm3)°$ with respect to crystal axes-a and -b. The gaps between stripe quasi-rectangular chains are very narrow (<50 nm), which presumably provides sufficiently strong connectivity to demonstrate bulk superconductivity in the sample. Similar behavior for the basal plane stripes was also observed in SEM images of $K_xFe_{1.6+y}Se_2$ single crystal surface [38]. Using the 3D mapping results, the average concentration of superconducting phase in our sample can be estimated as $16.6 \pm 1.1\%$. It is worth noting that in separate 2D cross-sections of Fig. 1(a)-(c) the stripes may sometimes seem isolated or terminated but the whole 3D stack shows that the superconducting paths are fully percolative in 3D. This is demonstrated in the volume view of superconducting phase in a small block of the sample (Fig. 1(d)).

### B. Ultrafast dynamics in the superconducting state

Alkali metal elements react readily with oxygen and water in air. Consequently, there have been few studies of the optical characteristics of alkaline iron selenide superconductors. In this paper we present results of dual-color pump-probe spectroscopy on $(Na_{0.32}K_{0.68})_{0.95}Fe_{1.75}Se_2$ single crystals demonstrating importance of SC/AFM phases coupling. We first discuss the results in the superconducting state in this subsection. Figure 2(a) shows a typical time and temperature evolution of $\Delta R/R$ obtained from $(Na_{0.32}K_{0.68})_{0.95}Fe_{1.75}Se_2$ single crystals. Three distinct regions separated by two characteristic temperatures at $T^*$ (~100 K) and $T_c$ (~30 K) are clearly evident from Fig. 2(a). At temperatures higher than 100 K, the negative $\Delta R/R$ responses decrease with increasing delay time and the period of the oscillation increases. At temperatures lower than 100 K, the Δ$R/R$ for the slow relaxation process shows a sign reversal from negative to positive. In the superconducting state, there is a signature of positive Δ$R/R$ until the lowest temperature of 20 K is reached and the amplitude of Δ$R/R$ increases when $T < T_c$ [Figs. 2(b) and 2(c)]. The pump–probe measurements generally show a sub-picosecond timescale drop for Δ$R/R$ at zero time delay, which is caused by the pump pulse-excited carriers (i.e., changes in the charge distribution), as shown in Fig. 3(a). Subsequently, the photoexcited quasiparticles (QPs) release their energy to phonon and spin subsystems. These chain reactions are clearly seen in the Δ$R/R$ measurements in Fig. 3(a).

The temperature-dependent Δ$R/R$ in Fig. 2(a) and Fig. 3(a) is shown as it is without any fitting process. The peak value for Δ$R/R$ (defined as $|A|$) is the response of the electron distribution near the Fermi surface (FS) [28,39,40]. Three temperature regions are clearly observed in Fig. 3(b). Namely, $|A|$ decreases as the temperature decreases from 290 K to 100 K, remains constant (about $8.1 \times 10^{-5}$) at temperatures lower than 100 K [see yellow dashed lines in Figs. 3(b) and 3(c)] and then rises dramatically when $T < T_c$. The metallic phase becomes superconducting when $T < T_c$ and there is an additional increase ($|A|$) in $|A|$ at temperatures lower than $T_c$, which is entirely due to the SC phase. Since $|A|$ is proportional to the number ($N$) of photoexcited QPs in one phase and $N$ is also proportional to the volume of that phase ($V$), the volume fraction for the superconducting phase could be estimated by $V_s / V_{total} = |A|(20\ K) / |A| = [(9.4 \times 10^{-5}) - (8.1 \times 10^{-5})] / (8.1 \times 10^{-5}) = 16\%$, where $V_{total}$ is the total volume of the probe area and $|A|$ is calculated using the average of $|A|$ from 40 K to 30 K [see Fig. 3(c)]. This agrees well with the concentration of superconducting phase *directly* found by 3D mapping technique in Fig. 1. Additionally, our result is also consistent with earlier studies of alkali metal iron selenide superconductors, which estimated the volume fraction for the minority SC phase at about 10%~20% [18,22,24].

When the temperature is lower than $T_c$, $|A|$ rises dramatically, as shown in Figs. 3(b) and 3(c), because of the opening of SC gap, which is a typical response reflecting the bottleneck effect due to gap opening [25,28,41,42]. Assuming that the temperature-dependent SC gap Δ($T$) obeys the BCS theory Δ($T$) = Δ(0)[1-($T/T_c$)]$^{1/2}$ and the density of the thermally-excited QPs ($n_T$) is defined by $n_T \propto [\Delta(T)T]^{1/2} \exp(-\Delta(T)/T)$ [43], then $|A|$ can be fitted using the relationship, $n_T \propto [|A(T \rightarrow 0)|/|A(T)|]-1$ [41,42], as shown in the inset of Fig. 3(c). The fits yield Δ(0)=10.8 meV, which is consistent with the value obtained using multiple Andreev reflections effect spectroscopy measurements and ARPES. [44,45].

### C. Ultrafast dynamics in the normal state

The curve for $|A|$ in Fig. 3(b) shows another characteristic temperature at 100 K.

Shown as an example in the inset of Fig. 3(a), the relaxation processes ($t > 0$) of $\Delta R/R$ for $(Na_{0.32}K_{0.68})_{0.95}Fe_{1.75}Se_2$ single crystals can be described by [31,32,40]:

$$\frac{\Delta R}{R} = A_e e^{-\frac{t}{\tau_e}} + A_p e^{-\frac{t}{\tau_p}} + A_s e^{-\frac{t}{\tau_s}} \qquad (1)$$

where the first term on the right-hand side represents the energy relaxation process for photoexcited electrons via electron-electron coupling, and $A_e$ is proportional to the population of the photoexcited electrons that participate in the electron-electron coupling process with a corresponding relaxation time $\tau_e$ on a sub-picosecond timescale (<1 ps), as shown in Fig. 4(a). The second and third exponential-decay components are associated with the energy relaxations of high-energy phonons and hot spins [42,46–48], with corresponding relaxation times of $\tau_p$ (~ 2 ps) and $\tau_s$ (>10 ps), in which the values of $A_p$ and $A_s$ are the respective amplitudes of these relaxation processes. To describe the significant oscillation in $\Delta R/R$, an additional term of $A_{LA} e^{-\frac{t}{\tau_{LA}}} \cos\left(\frac{2\pi t}{P} + \phi\right)$ is necessary to add into Eq. (1) with the amplitude $A_{LA}$, damping time $\tau_{LA}$, oscillation period $P$ and initial phase of oscillation $\phi$.

The temperature-dependent amplitudes ($A_{LA}$) and damping time ($\tau_{LA}$) for the oscillation component in Fig. 4(b) are associated with the energy transfer in crystals via the propagation of strain pulses (LA phonons) [31,34,49]. According to Thomsen's model [50], the oscillation period $P$ of an oscillating component is expressed as $P = \lambda_{probe}/2nv_s$, where $\lambda_{probe}$ (=800 nm) is the wavelength of the probe beam, $n$ is the real part of the refractive index and the strain pulse propagates in crystals with velocity $v_s$. The propagating velocity $v_s$ at room temperature is estimated to be 2.98 km/s, using $n$=2.4 [38] and $P$=55.9 ps. The penetration depth for the strain pulse at room temperature is $l_{LA} = v_s \tau_{LA} = (2.98\ \text{km/s})(18.5\ \text{ps}) = 55.1$ nm, which, in principle, must be less than the skin depth of the probe beam ($l_s$). If it is assumed that a $(Na_{0.32}K_{0.68})_{0.95}Fe_{1.75}Se_2$ crystal is 100% metal, the skin depth ($l_s$) is about 35.4 nm (evaluated using the skin depth of an electromagnetic wave in metal, $\lambda_{probe}/4\pi k$, where $k$ is the imaginary part of the refractive index, $k$=1.8 [38]). However, the penetration depth that is obtained from the pump-probe measurements is more than 35.4 nm. Therefore, the oscillation component of $\Delta R/R$ is primarily generated by the photoexcited responses in the region of the insulating AFM phase, rather than the metallic phase.

In addition to the characteristics at $T_c$ and 100 K seen in Fig. 3(b), an additional significant change in the amplitudes ($A_{LA}$) and damping time ($\tau_{LA}$) for the oscillation component at 230 K is clearly observed in Figs. 4(a) and 4(b), as well. In general, the oscillation component of $\Delta R/R$ represents the propagation of energy in the lattice by phonons. Therefore, the oscillatory feature is expected to change with the appearance of the structural phase transition [31]. Consequently, the anomalous changes in $A_{LA}$ and $\tau_{LA}$ imply that a structural phase transition is taking place at ~230 K. Recently, this structure phase transition has been observed for alkali metal iron selenide superconductors at < 300 K using spatial-resolved micro x-ray diffraction and was associated with the orbital selective Mott phase (OSMP) transition [22]. The OSMP has been shown to connect the electronic band structure and the characteristic of spatially separated phases if there is a strong correlation between electrons [20–22]. At temperatures higher than the OSMP transition temperature $T_H$ (see Fig. 4(c)), the electrons in the $d_{xz}/d_{yz}$ bands are itinerant and the electrons in the $d_{xy}$ band are Mott localized because there is a gap opening in the $d_{xy}$ band. Therefore, itinerant and Mott localized electrons coexist in alkali metal iron selenide superconductors when $T > T_H$. When the temperature is lower than $T_H$, the gap in the $d_{xy}$ band vanishes and all of the 3$d$ bands become metallic [23], which produces a hump in the resistivity measurement, as shown in Fig. 4(c) [21]. A metallic interface phase also appears between the metallic and AFM phase regions because of OSMP transition and reduces the volume fraction for the AFM region when $T < T_H$ [22].

However, the metallic interface phase is not uniform throughout the entire crystal. The area of the AFM phase is decreased by the formation of the metallic interface phase, in which the lattice slightly elongates (~0.2 Å) along the $c$-axis [22]. When the temperature deceases, the interface phase region expands and, in turn, damps the propagation of strain pulses in the crystals, leading to a decrease of damping time for the strain pulses $\tau_{LA}$ when $T < T_H$. Nevertheless, the amplitude ($A_{LA}$) of the oscillation component increases when the temperature is lower than $T_H$, as shown in a previous study of K$_x$Fe$_{2-y}$Se$_2$, which is attributed to the disappearance of a gap in the $d_{xy}$ band [34]. A gap only opens in the $d_{xy}$ band when the temperature is higher than $T_H$. The opening of this gap at $T > T_H$ allows additional relaxation processes for photoexcited electrons, such as interband recombination or intraband scattering. Therefore, less energy is transferred from the photoexcited electrons to phonons, which drives the strain pulses, i.e., the value of $A_{LA}$ decreases when $T > T_H$. However, more energy is transferred to the strain pulses when the gap closes at $T < T_H$, so the value of $A_{LA}$ increases as the temperature decreases.

### D. The ultrafast coupling channel between the metallic phase and the AFM phase

The photoexcited electrons release their energy via electron-electron coupling, with a relaxation time of $\tau_e$. This relaxation process can occur either in the metallic phase or in the AFM phase, with corresponding respective relaxation times of $\tau_{e,m}$ and $\tau_{e,AFM}$, as shown in Fig. 5. In this study, the total responses of $\Delta R/R$ must come from both the metallic and the AFM phases, since the diameter of the probe area is about 60 μm, which is larger than the domain sizes (1~20 μm, larger domain sizes dominate at low temperatures [15,18,22]) for the two phases. If there is no interaction between the two main phases, the final relaxation time $\tau_e$ should be a linear combination of two parallel relaxation rates prevailing in different phases, i.e., the inverse photoexcited electron relaxation time $1/\tau_e$ is the linear sum of $V_m(1/\tau_{e,m})$ and $V_{AFM}(1/\tau_{e,AFM})$, where $V_m$ ~16 % and $V_{AFM}$ ~84% are the respective volume fractions for the metallic phase and the AFM phase [18, 24]. When the temperature is lower than $T_H$, $V_{AFM}$ decreases with decreasing the temperature because of the expansion of the metallic interface phase region. Therefore, the final relaxation time $\tau_e$ gradually increases when the temperature is lower than $T_H$ [see the red-dashed line in Fig. 4(a)]. However, the value of $\tau_e$ below 100 K ($T^*$) shows a significant deviation from the red-dashed line and saturates below $T_c$. One possible scenario for this intriguing novel phenomenon is the existence of an additional interaction channel between the metallic and the AFM phases at temperatures lower than $T_H$. If the photoexcited electrons in two phases exchange energy with each other and the coupling strength (which is proportional to the relaxation rate of $1/\tau_{e,c}$) increases when the temperature decreases, the temperature-dependent $\tau_e$ in Fig. 4(a) is described by the red-solid line below $T^*$. Consequently, the clear electronic correlation between the two main phases further enhances the energy relaxation for electrons in high-energy states and the value of $\tau_e$ decreases. This additional interaction channel between the metallic and the AFM phases is turned on at 100 K, which may lead to the orbital-selective pairing proposed recently [19].

## IV. CONCLUSIONS

Nano-scale phase-separation in $(Na_{0.32}K_{0.68})_{0.95}Fe_{1.75}Se_2$ single crystals is explored using dual-color pump-probe spectroscopy and hybrid FIB-SEM system. *Both 3D FIB-SEM mapping and ultrafast spectroscopy consistently reveal a 16% volume fraction of the nano-superconducting phase. The latter method further suggests a superconducting gap of 10.8 meV at 20 K. The sound velocity of 2.98 km/s is calculated using the period of the oscillatory feature on $\Delta R/R$. Moreover, abnormal changes in the temperature dependence of the oscillatory feature and the QP relaxation process are observed at $T_H$ =230 K, which is consistent with the onset temperature of OSMP transition identified in $K_xFe_{2-y}Se_2$ superconductors, indicative of the appearance of the metallic interface phase in $(Na_{0.32}K_{0.68})_{0.95}Fe_{1.75}Se_2$ single crystals. Intriguingly, the interaction channel between the metallic and AFM phases opens at the temperature lower than $T_H$. The coupling between the nano-metallic and the AFM phases through the metallic interface phase is established at temperature as low as $T^*$=100 K. QP ultrafast dynamics also shows the energy transfer channel between these two separated phases, which is crucial to superconductivity and provides new insight for the pairing mechanism in alkali metal iron selenide superconductors.

## ACKNOWLEDGMENTS

This work was supported by the Ministry of Science and Technology of the Republic of China, Taiwan (Grant No's. 107-2119-M-009-010-MY2, 106-2119-M-009-013-FS and 106-2628-M-009-003-MY3). This work was also financially supported by the Center for Emergent Functional Matter Science of National Chiao Tung University from The Featured Areas Research Center Program and the Research Team of Photonic Technologies and Intelligent Systems at NCTU within the framework of the Higher Education Sprout Project by the Ministry of Education (MOE) in Taiwan. This work has been supported by the Ministry of Education and Science of the Russian Federation in the framework of Increase Competitiveness Program of NUST "MISiS" Grant No. K2-2017-084; by Act 211 of the Government of Russian Federation, Contracts No. 02.A03.21.0004 and No. 02.A03.21.0011. AM and AB thank the support RSF, grant No. 16-42-01100. AAZ and JW acknowledge the support of the EU Graphene Flagship Project and European Research Council Synergy Grant Hetero2D.

**Figure captions**

FIG. 1. Morphology of the SC-AFM phase separation represented by three cross sections for main $(Na_{0.32}K_{0.68})_{0.95}Fe_{1.75}Se_2$ crystal planes. Superconducting phase has a brighter contrast. Yellow scale bar is equal to 1 μm. Direction of image x- and y-axes corresponds to (a) a & c, (b) b & c and (c) a & b, respectively. Volume view for superconducting phase in a cubic shape block (side length 1.625 μm) is shown in (d) These results of FIB-SEM nanotomography are obtained using SEM pixel size of 6.77 nm and FIB milled slice thickness 9 nm. The size of the 3D stack is 4.644 × 2.823 × 3.845 μm³ along a-, b- and c-axes, respectively.

FIG. 2. The relationship between temperature and delay time for $\Delta R/R$ in a $(Na_{0.32}K_{0.68})_{0.95}Fe_{1.75}Se_2$ single crystal. (a) The solid lines indicate $\Delta R/R = 0$. (b) A magnified representation of a part of the $\Delta R/R$ in (a). (c) Temperature dependence for zero field cooled (ZFC) and field cooled (FC) magnetic moment $M(T)$. The dashed lines respectively represent the superconducting transition temperature ($T_c$) and 100 K ($T^*$).

FIG. 3. Typical transient reflectivity changes ($\Delta R/R$) and quantitative analysis at various temperatures. (a) The red-solid line in the inset is the fitting curve using Eq. (1). (b) The temperature dependence for |$A$| (defined as the amplitude of $\Delta R/R$, |$A$|, at zero delay time). (c) A magnification of a part of the temperature-dependent |$A$| in (b). The inset shows the density of thermally-excited QPs, $n_T(T)$. The red-solid lines are the fitting of |$A$| and $n_T(T)$ at temperatures less than $T_c$. The yellow-dashed lines are a guide to the eyes.

FIG. 4. Temperature-dependent relaxation processes of photoexcited QPs and acoustic phonons. (a) The temperature dependence of the amplitude |$A_e$| and the relaxation time $\tau_e$ for the relaxation process of photoexcited QPs, by fitting Eq. (1). (b) The temperature dependence of the oscillation amplitude $A_{LA}$ and the damping time $\tau_{LA}$ for the oscillation component, by fitting Eq. (1). (c) The temperature dependence for the resistivity $\rho$ and its slope for a $(Na_zK_{1-z})_xFe_{2-y}Se_2$ single crystal. The red-solid and -dashed lines in (a) represent the linear sum of the parallel relaxation rates for the metallic phase ($1/\tau_{e,m}$) and the AFM phase ($1/\tau_{e,AFM}$), with (solid) and without (dashed) electronic coupling between the two phases using a time constant $\tau_{e,c}$. The blue- and black-solid lines are a guide to the eyes.

FIG. 5. A schematic representation of the relaxation and energy transfer for photoexcited electrons in phase-separated $(Na_{0.32}K_{0.68})_{0.95}Fe_{1.75}Se_2$ signal crystals.

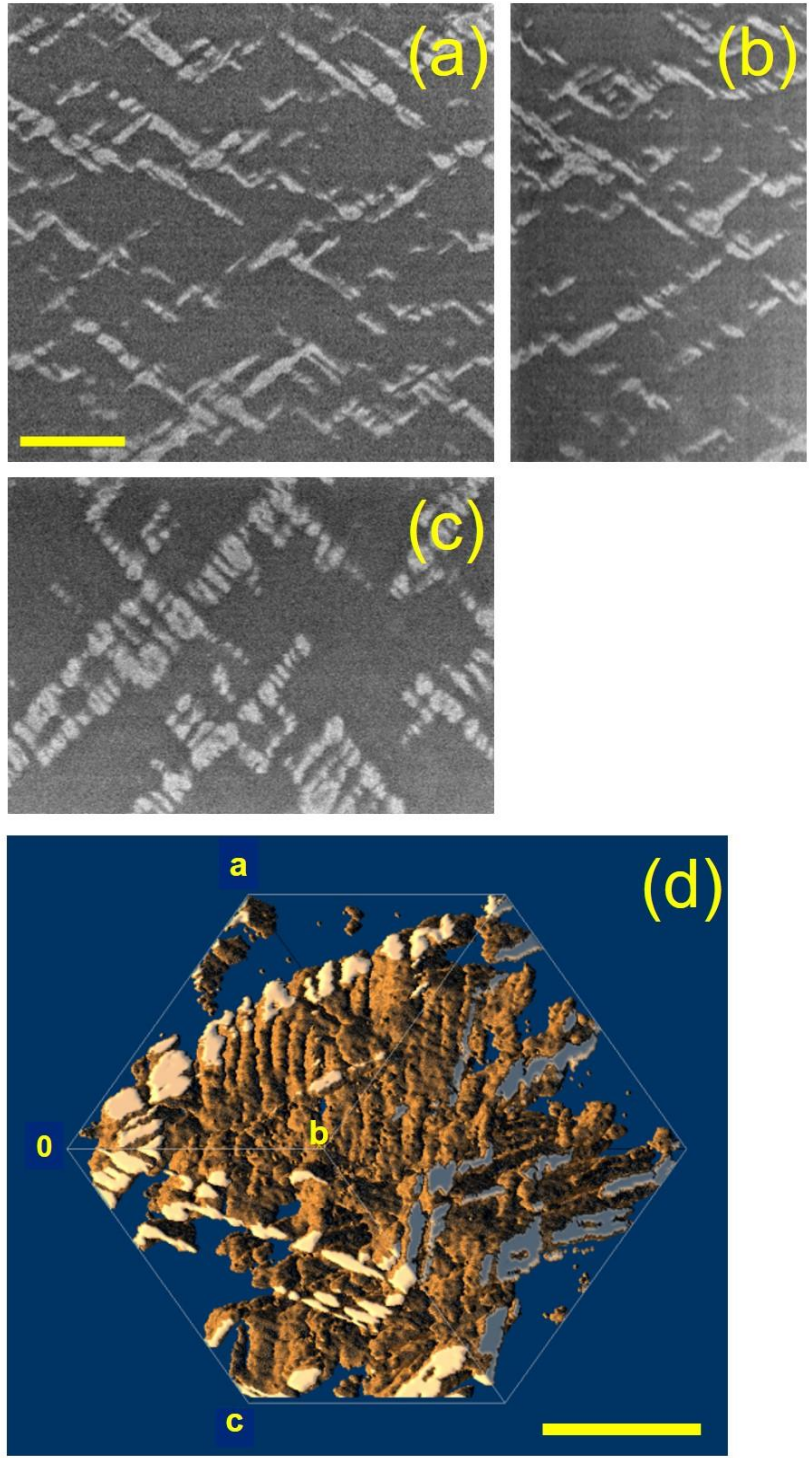

**FIG. 1**

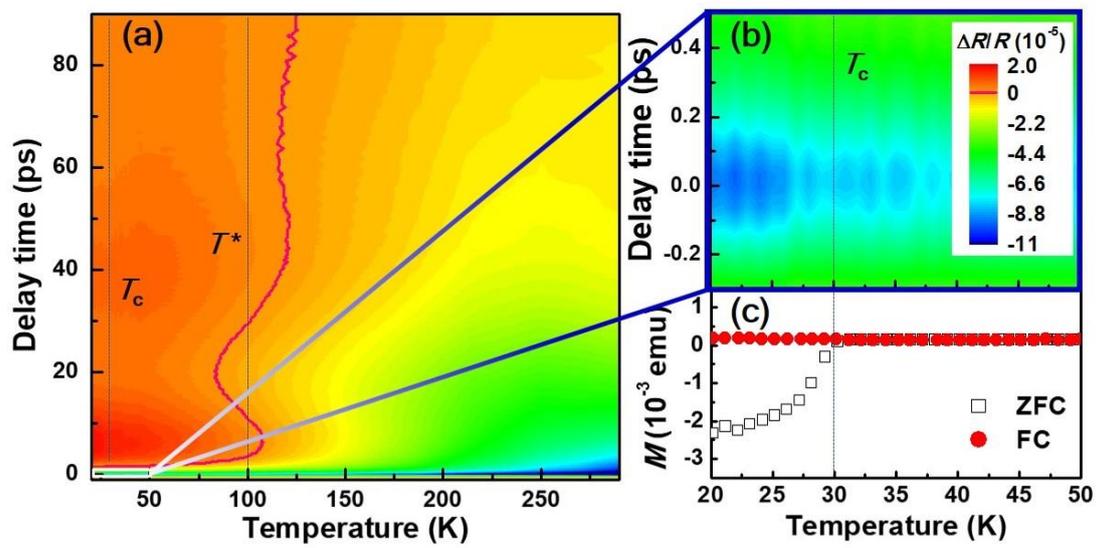

**FIG. 2**

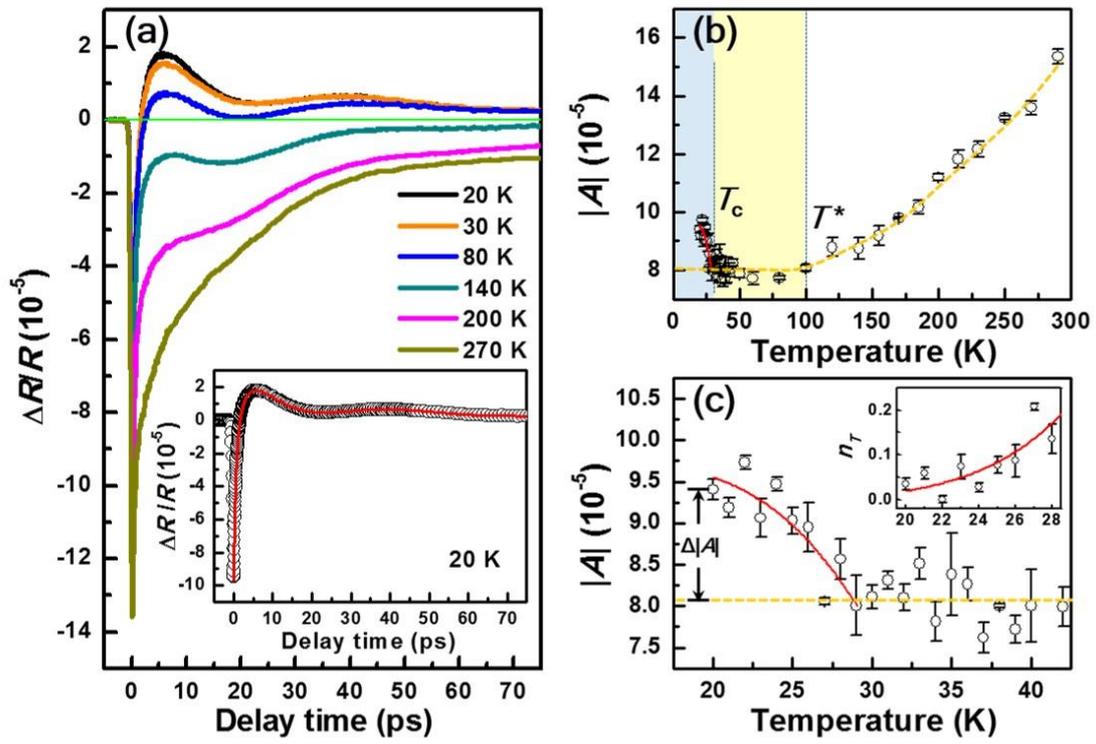

FIG. 3

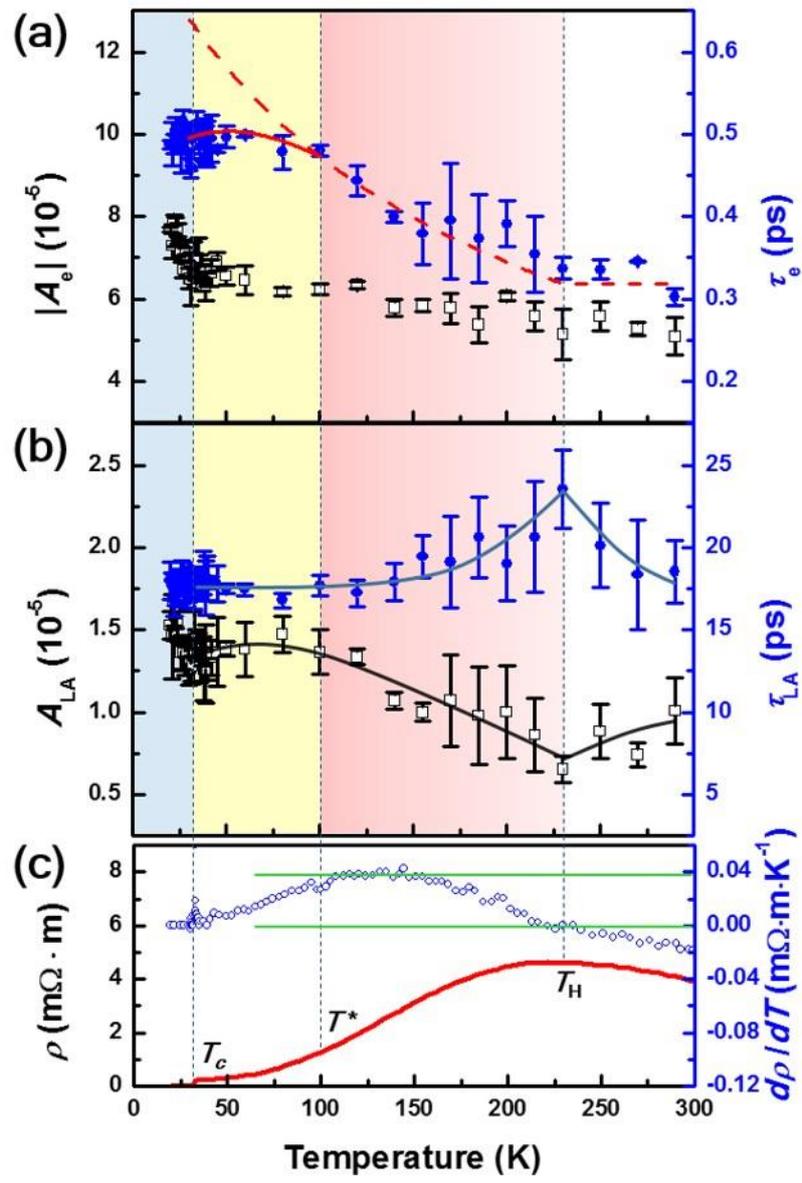

FIG. 4

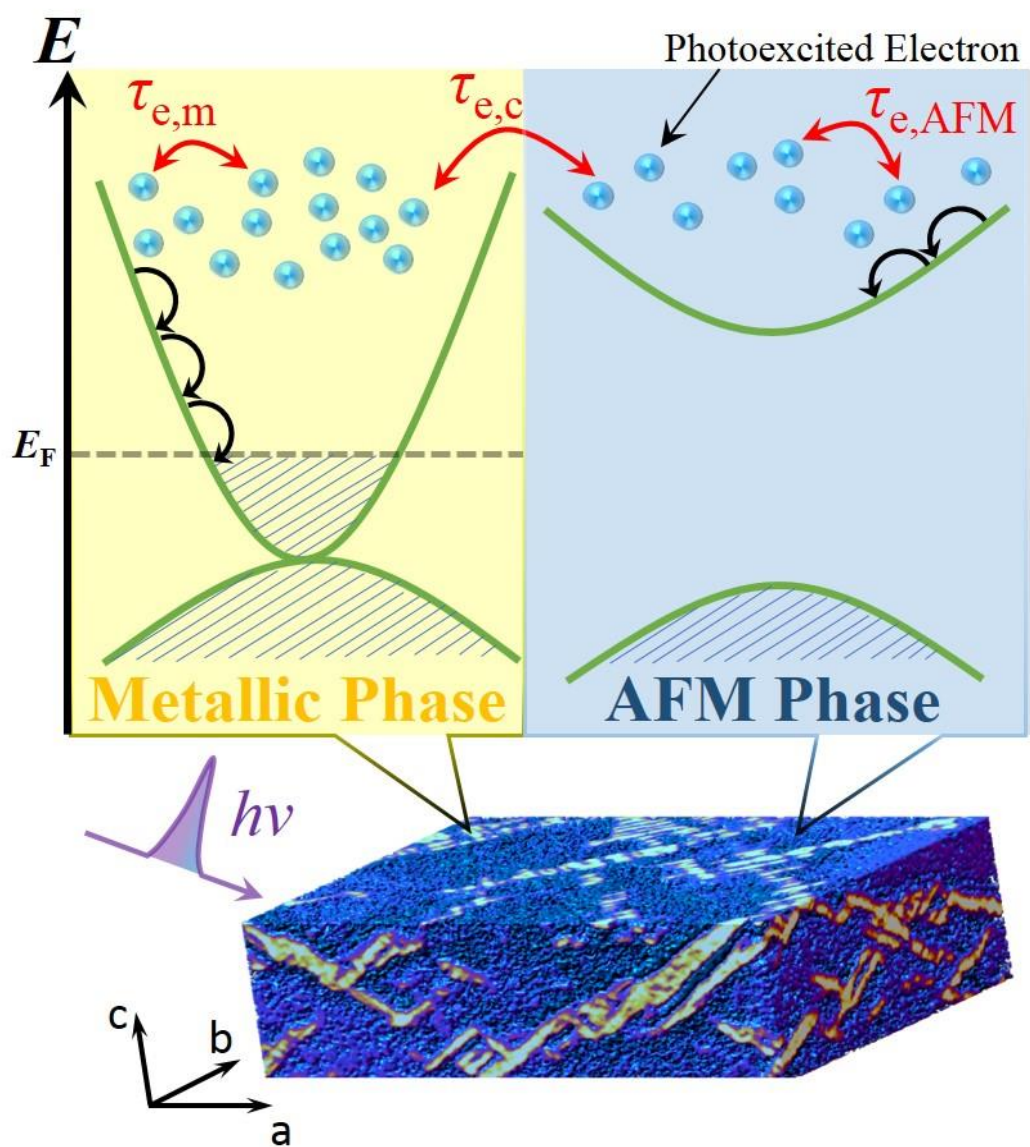

$(Na_{0.32}K_{0.68})_{0.95}Fe_{1.75}Se_2$ single crystal

**FIG. 5**